\documentclass[aps,prl,showpacs,twocolumn,groupedaddress]{revtex4}
\usepackage{graphicx}

\begin{document}

\title{Chiral magnetic effect in lattice QCD with a chiral chemical potential}

\author{Arata~Yamamoto}
\affiliation{Department of Physics, The University of Tokyo, Tokyo 113-0033, Japan}

\date{\today}

\begin{abstract}
We perform a first lattice QCD simulation including two-flavor dynamical fermion with a chiral chemical potential.
Because the chiral chemical potential gives rise to no sign problem, we can exactly analyze a chirally imbalanced QCD matter by the Monte Carlo simulation.
By applying an external magnetic field to this system, we obtain a finite induced current along the magnetic field, which corresponds to the chiral magnetic effect.
The obtained induced current is proportional to the magnetic field and to the chiral chemical potential, which is consistent with an analytical prediction.
\end{abstract}


\maketitle

Quantum chromodynamics (QCD) is expected to have many characteristic phases at finite baryon density.
However, it is difficult to extract exact information about high-density QCD.
One major reason is that lattice QCD simulation suffers from the sign problem.
At finite baryon chemical potential, the fermion determinant becomes complex, and its phase fluctuation makes it severe to evaluate the ensemble average in the Monte Carlo simulation.
It is instructive to study the exceptional cases which can avoid the sign problem in lattice QCD.
Through these studies, we can learn the qualitative behavior of high-density QCD.
The well-studied examples are isospin chemical potential and two-color QCD.
We here consider another exceptional case, that is, {\it chiral chemical potential} \cite{Fukushima:2008xe}.

The chiral chemical potential $\mu_5$ is defined in the fermion part of the Euclidean action as
\begin{eqnarray}
S_F = \int d^4x \bar{\psi} ( \gamma_\mu D_\mu + m + \mu_5 \gamma_4 \gamma_5 ) \psi \ .
\label{eqaction}
\end{eqnarray}
This action preserves positive semi-definite property of the fermion determinant and thus has no sign problem.
Physically, the chiral chemical potential generates an imbalance between the right-handed and left-handed fermions.
From the index theorem, this imbalance is equivalent to the topological charge of the background gauge field.
The chiral chemical potential is related to the space-time dependent effective theta angle, which represents local variation of the topological charge \cite{Fukushima:2008xe}.
The constant chiral chemical potential is regarded as a static alternative to the topology changing effect.
Since lattice QCD can simulate only an imaginary-time equilibrium system not a real-time evolution, the chiral chemical potential is a reasonable choice to study this effect in lattice QCD.

One of the most important applications of the chiral chemical potential is the chiral magnetic effect.
The chiral magnetic effect is an electromagnetic charged current perpendicular to the reaction plane of a heavy-ion collision \cite{Kharzeev:2004ey}.
A noncentral collision of two heavy ions produces a strong magnetic field, and then this magnetic field fixes the spin and momentum directions of the quarks according to their chiralities. 
As a consequence, a finite net current is induced along the magnetic field if there is a chiral imbalance associated with the axial anomaly or, equivalently, a nontrivial topology of the background gluon configuration.
The research for the chiral magnetic effect is important because it enables us to detect topological structure and local $\mathcal{P}$ and $\mathcal{CP}$ violation in the strong interaction by experiments \cite{:2009uh}.
The chiral chemical potential has been introduced in several phenomenological works for the chiral magnetic effect \cite{Fukushima:2008xe,Kharzeev:2009pj,Fukushima:2010fe,Fukushima:2010zza}.
The chiral magnetic effect has also been studied in lattice gauge theory without the chiral chemical potential \cite{Buividovich:2009wi,Abramczyk:2009gb,Braguta:2010ej}.

In this work, we performed a first lattice QCD simulation with the chiral chemical potential.
For the lattice gauge action, we used the plaquette gauge action with $N_c=3$.
For the lattice fermion action, we used the Wilson-Dirac operator as
\begin{eqnarray}
D(\mu_5)_{x,y} &=& \delta_{x,y}
- \kappa \sum_i \Bigl[ (1-\gamma_i)U_i(x) \delta_{x+\hat{i},y} \nonumber \\
&&+ (1+\gamma_i)U^\dagger_i(x-\hat{i}) \delta_{x-\hat{i},y} \Bigr] \nonumber\\
&&- \kappa \Bigl[ (1-\gamma_4 e^{a\mu_5\gamma_5})U_4(x) \delta_{x+\hat{4},y} \nonumber\\
&&+ (1+\gamma_4 e^{-a\mu_5\gamma_5})U^\dagger_4(x-\hat{4}) \delta_{x-\hat{4},y} \Bigr] \ .
\label{eqDirac}
\end{eqnarray}
The chiral chemical potential is introduced as the exponential matrix factor $e^{\pm a\mu_5\gamma_5}=\cosh (a\mu_5) \pm \gamma_5\sinh (a\mu_5)$, which is analogous to baryon chemical potential on the lattice \cite{Hasenfratz:1983ba}.
The lattice action reproduces the original action (\ref{eqaction}) in the continuum limit $a\to 0$.
This Dirac operator satisfies the relation $\gamma_5 D(\mu_5) \gamma_5 =  D^\dagger (\mu_5)$ even at $a\ne 0$, and thus the fermion determinant $\det D(\mu_5)$ is always positive real.
We generated the gauge configurations with the $N_f=2$ dynamical Wilson fermion by the Hybrid Monte Carlo algorithm.
The inversion of the Dirac operator was calculated by the BiCGstab solver with the even-odd preconditioning.
In most part of the following analyses, the lattice gauge coupling is fixed at $\beta=2N_c/g^2=5.32144$, and the hopping parameter is fixed at $\kappa=0.1665$ both for the valence and dynamical fermions.
These values correspond to the lattice spacing $a \simeq 0.13$ fm ($a^{-1} \simeq 1.5$ GeV) and the pion mass $am_\pi \simeq 0.26$ ($m_\pi \simeq 0.4$ GeV) at $\mu_5 = T = 0$ \cite{Orth:2005kq}.

\begin{figure}[t]
\begin{center}
\includegraphics[scale=1.2]{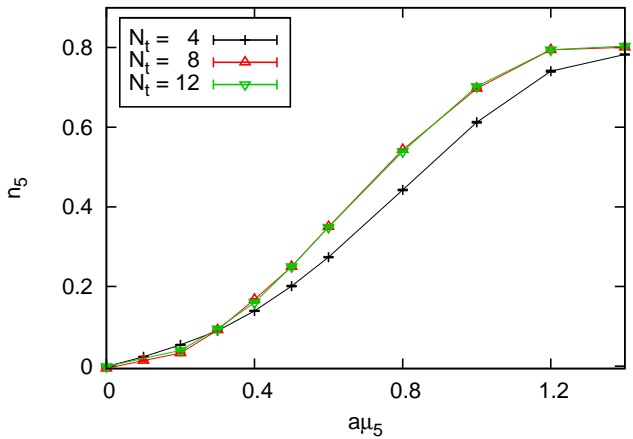}
\caption{\label{fig1}
The chiral charge density $n_5$.
The lattice sizes are $N_s^3 \times N_t = 12^3\times 4, 12^3\times 8$, and $12^4$.
}
\end{center}
\end{figure}

First, we calculated the chiral charge density
\begin{eqnarray}
n_5 \equiv -a^3\langle \bar{\psi} \gamma_4 \gamma_5 \psi \rangle
= a^3\langle \psi^\dagger_R \psi_R - \psi^\dagger_L \psi_L \rangle \ ,
\end{eqnarray}
i.e., the difference between the particle number densities of the right-handed and left-handed fermions.
The numerical results are shown in Fig.~\ref{fig1}.
The chiral charge density is exactly zero at $\mu_5 = 0$, where the right-handed and left-handed fermions are symmetric.
The chiral charge density becomes nonzero in $\mu_5>0$ and increases as $\mu_5$ increases.
Note, however, that the chiral charge density saturates around $a\mu_5= 1.4$.
This is due to the saturation of the lattice sites, which is observed also in the cases of isospin chemical potential and two-color QCD \cite{Kogut:2002tm}.
Because the saturation is a lattice artifact, we cannot trust the data in $a\mu_5> 1.0$.

Figure \ref{fig1} includes the data with different values of temperature $T = 1/(aN_t)$.
The low-temperature ones ($N_t=8$ and 12) are in the confinement phase, and their behaviors are almost the same in the present calculation.
The high-temperature one ($N_t=4$) is in the deconfinement phase.
The chiral charge density is qualitatively different between the confinement and deconfinement phases.
In $a\mu_5 \le 0.2$, the chiral charge density in the deconfinement phase is larger than that in the confinement phase, which is consistent with phenomenological models \cite{Fukushima:2010fe,Chernodub:2011fr,Ruggieri:2011xc}.
In $a\mu_5 > 0.2$, the chiral charge density in the confinement phase grows rapidly, and becomes larger than that in the deconfinement phase.
This rapid growth is considered to be caused by some bound-state contribution.

Next, we focus on the phase structure of a QCD matter in $\mu_5 > 0$.
In two-flavor QCD at $\mu_5 = 0$, the confinement/deconfinement phase transition is a crossover in the non-chiral limit (and a second-order phase transition in the chiral limit).
There is a possibility that the order of the transition is changed by introducing the chiral chemical potential.
For example, phenomenological models predict a first-order phase transition line in the $\mu_5$-$T$ plane \cite{Fukushima:2010fe,Chernodub:2011fr,Ruggieri:2011xc}.

\begin{figure}[t]
\begin{center}
\includegraphics[scale=1.2]{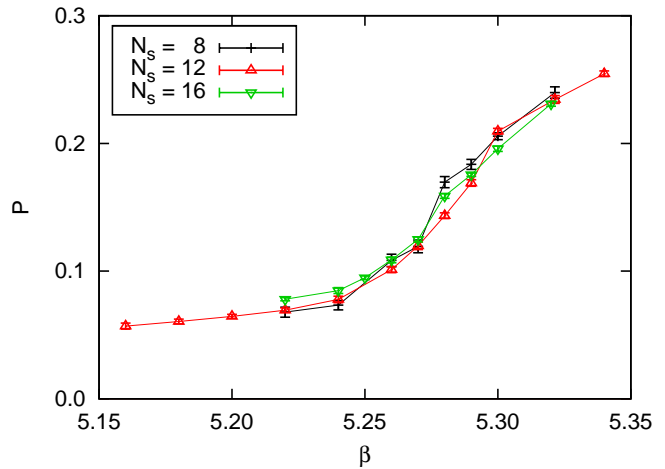}
\caption{\label{fig2}
The expectation value of the Polyakov loop $P$ at $a\mu_5 = 1.0$.
The lattice sizes are $N_s^3 \times N_t = 8^3\times 4, 12^3\times 4$, and $16^3\times 4$.
}
\end{center}
\end{figure}

We calculated the temperature dependence of the Polyakov loop, which is a good indicator of confinement/deconfinement, by varying the lattice gauge coupling $\beta$.
The numerical result at $a\mu_5 = 1.0$ is shown in Fig.~\ref{fig2}.
The Polyakov loop increases around $\beta \simeq 5.27$, which means a deconfinement transition driven by thermal effects.
However, the Polyakov loop and its susceptibility are almost independent of the spatial volume $V=a^3N_s^3$.
This scaling behavior indicates that this transition is not a true phase transition but a smooth crossover.
We cannot obtain a signal of a true phase transition in $a\mu_5 \le 1.0$.
Therefore, we conclude that, even if it exists, there is no change of the order of the phase transition in the present accessible range of $\mu_5$.
This difficulty has been already experienced in lattice QCD with a finite isospin chemical potential $\mu_I$ \cite{Kogut:2002tm}.
Although the critical endpoint is located away from $\mu_I = 0$ when the quark mass is not quite small.
Since the location of the critical endpoint depends on the quark mass, the situation will change in the near-chiral limit.

As shown above, we obtained a QCD matter with an imbalance between the right-handed and left-handed fermion number densities.
To analyze the chiral magnetic effect, we applied an external magnetic field to this system.
On the lattice, the U(1) electromagnetic gauge field is introduced as the Abelian phase factor $u_\mu(x)$ on the SU($N_c$) link variable $U_\mu(x)$ in the Dirac operator (\ref{eqDirac}).
For generating a constant magnetic field $B$ in the $x_3$-direction, the phase factor is taken to be $u_2(x)=\exp(iaqBx_1)$, $u_1(x_1=aN_s)=\exp(-iaqBN_sx_2)$, and $u_\mu(x)=1$ for other components \cite{Buividovich:2009wi}.
Only discrete values of the magnetic field are allowed as $a^2qB=(2\pi/N_s^2)\times$(integer) because of the quantization of the magnetic flux.
Since the magnetic field is not dynamical but external, there is no backreaction from the quarks to the electromagnetic gauge field.
For simplicity, we consider two fermion flavors with the same charge $q$.
This approximation does not change the qualitative behavior since the charge difference between the quarks is not essential for the underlying mechanism of the chiral magnetic effect.

\begin{figure}[t]
\begin{center}
\includegraphics[scale=1.2]{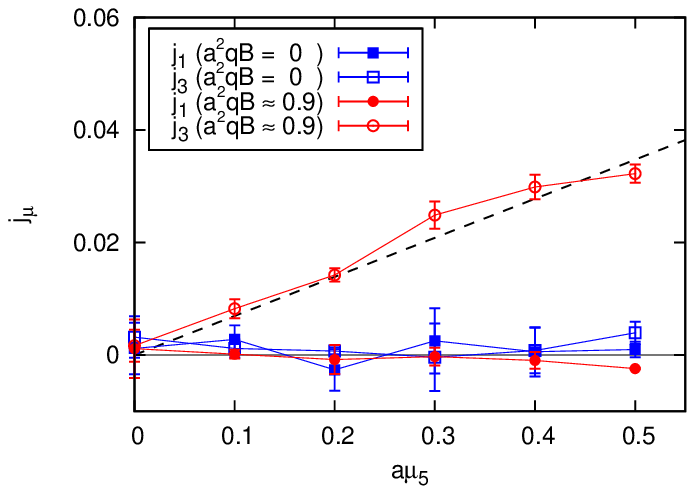}
\caption{\label{fig3}
The transverse current density $j_1$ and the longitudinal current density $j_3$ as a function of the chiral chemical potential $\mu_5$.
The black dashed line is a linear function (\ref{eqj3}).
The lattice size is $N_s^3 \times N_t = 12^3\times 4$.
}
\end{center}

\begin{center}
\includegraphics[scale=1.2]{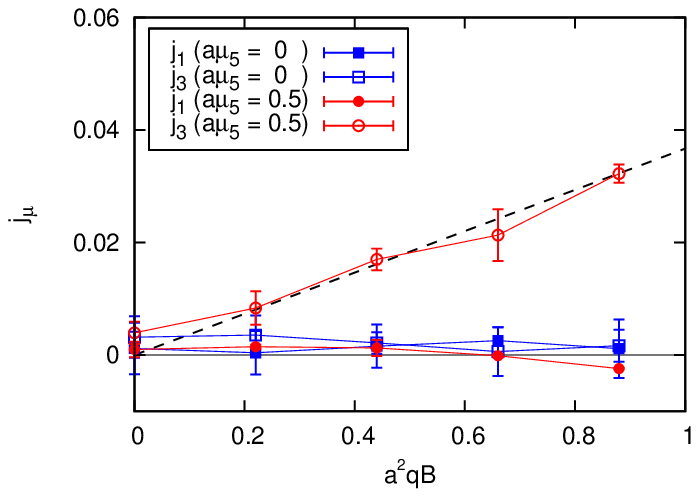}
\caption{\label{fig4}
The transverse current density $j_1$ and the longitudinal current density $j_3$ as a function of the magnetic field $B$.
}
\end{center}
\end{figure}

In this setup, we measured the vector current density
\begin{eqnarray}
j_\mu \equiv a^3\langle \bar{\psi} \gamma_\mu \psi \rangle \ .
\label{eqJL}
\end{eqnarray}
The simulations were done in the deconfinement phase ($N_t=4$), which is relevant for the chiral magnetic effect in heavy-ion collisions.
The transverse component $j_1$ and the longitudinal component $j_3$ of the current density are depicted as a function of $\mu_5$ in Fig.~\ref{fig3}, and as a function of $qB$ in Fig.~\ref{fig4}.
The two transverse components of the current density are the same, $j_1=j_2$, from the rotational symmetry, and they are zero in all of the simulations.
All components of the current density are zero either at $B=0$ or at $\mu_5=0$.
Only when both $B$ and $\mu_5$ are nonzero, a finite current density is generated in the longitudinal direction.
These results suggest that an external magnetic field induces a finite current density along the magnetic field only in a chirally imbalanced QCD matter. 
This is exactly what is expected of the chiral magnetic effect.

As seen in Figs.~\ref{fig3} and \ref{fig4}, the induced current density is an increasing function of $\mu_5$ and $qB$.
Furthermore, it is given as a linearly rising function both of $\mu_5$ and of $qB$.
We can parametrize its functional form as
\begin{eqnarray}
j_3 = a^3 C N_{\rm dof} \mu_5 qB \ .
\label{eqj3}
\end{eqnarray}
The factor $N_{\rm dof}$ is the number of particles with the same charge, which is 6 ($= N_c \times N_f$) in this simulation.
The overall constant is numerically determined as $C= 0.013 \pm 0.001$ by fitting the data.
This functional form has been predicted by an analytical approach using the Dirac equation coupled with the background magnetic field \cite{Fukushima:2008xe}.
The lattice result establishes this prediction, except for the overall constant $C$, which is $1/(2\pi^2) \simeq 0.05$ in the analytical approach.
This deviation comes from some QCD corrections.
One possible candidate is a correction by the renormalization.
The local vector current (\ref{eqJL}) is not renormalization-group invariant on the lattice \cite{Karsten:1980wd}.
This is very different from the vector current in the continuum.
Another candidate is the dielectric correction, which reduces the induced current \cite{Fukushima:2010zza}.

The above situation is completely different from the standard lattice QCD without the chiral chemical potential.
In the standard lattice QCD, we cannot observe the global induced current.
Because the current itself is zero and only its local fluctuation is nonzero, the chiral magnetic effect is studied only through the local fluctuation \cite{Buividovich:2009wi}.
In principle, lattice QCD can reproduce the gauge configuration with a nontrivial topology, which gives a finite chiral imbalance via the index theorem.
However, the global topological charge (or the global chiral charge) per volume is negligibly small, unless one artificially makes the gauge configuration with a huge number of topological charge.
The magnetic field cannot induce the global current in observable amount.
On the other hand, at a finite chiral chemical potential, the chiral charge density is finite and independent of the volume.
Therefore, owing to the introduction of the chiral chemical potential, we can observe the global current induced by the chiral magnetic effect.

\begin{figure}[t]
\begin{center}
\includegraphics[scale=1.2]{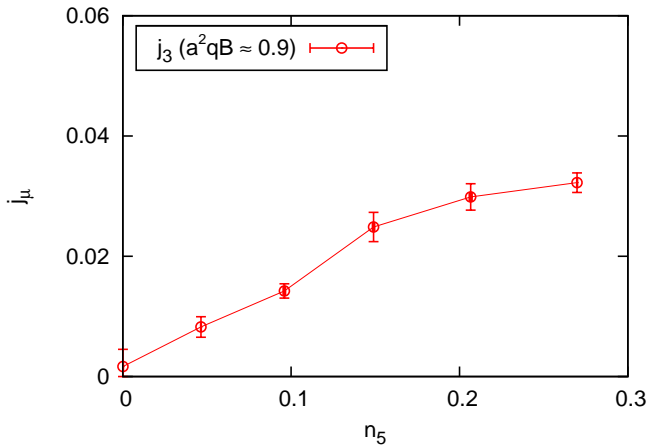}
\caption{\label{fig5}
The longitudinal current density $j_3$ as a function of the chiral charge density $n_5$.
}
\end{center}
\end{figure}

Finally, in Fig.~\ref{fig5}, we plot the induced current density as a function of the chiral charge density $n_5$ with a fixed magnetic field.
We can see that the induced current density is approximately proportional to $n_5$.
In this simulation, the magnetic field is very large, $qB \gg \mu_5^2$.
Under the strong magnetic field, the quantum state of a charged particle is dominated by the lowest Landau level.
While the induced current density cannot generally be written in a simple analytical function of $n_5$, the contribution of the lowest Landau level can be written as a linear function in the chiral limit \cite{Fukushima:2008xe}.
Although the lattice QCD simulation is not in the chiral limit, we can expect that the obtained induced current is dominated by the lowest Landau level.

In this study, we have performed the lattice QCD simulation with the chiral chemical potential, and verified that it actually works well.
We have succeeded in obtaining nonzero current induced by an external magnetic field, which is directly related to the chiral magnetic effect.
This is a great advantage because, otherwise, the chiral magnetic effect is indirectly analyzed through the local fluctuation of the current and the local fluctuation is easily affected by various contaminations. 
Although we have not addressed the dependence on the quark mass and the effect of renormalization, these issues can be studied systematically by the standard lattice QCD techniques.
It is also straightforward to improve details of the numerical simulation, such as, lattice action, lattice spacing, lattice volume, etc.

This work was supported in part by the Grant-in-Aid for Scientific Research in Japan under Grant No.~22340052.
The lattice QCD simulations were carried out on NEC SX-8R in Osaka University.

\end{document}